# Universal fluctuations: a new approach to the study of "phase transitions" in intermediate energy heavy ion collisions[+]

J. D. Frankland[1,*], R. Bougault[2], A. Chbihi[1], S. Hudan[1], A. Mignon[1], G. Auger[1], Ch. O. Bacri[3], N. Bellaize[2], B. Borderie[3], B. Bouriquet[1], A. Buta[2], J. L. Charvet[4], J. Colin[2], D. Cussol[2], R. Dayras[4], N. De Cesare[5], A. Demeyer[6], D. Doré[4], D. Durand[2], E. Galichet[3,7], E. Gerlic[6], D. Guinet[6], B. Guiot[1], G. Lanzalone[8], Ph. Lautesse[6], F. Lavaud[3], J. L. Laville[1], J. F. Lecolley[2], R. Legrain[4,#], N. Le Neindre[1], O. Lopez[2], L. Nalpas[4], J. Normand[2], M. Pârlog[9], P. Pawlowski[3], E. Plagnol[3], M. F. Rivet[3], E. Rosato[5], R. Roy[10], G. Tabacaru[9], B. Tamain[2], E. Vient[2], M. Vigilante[5], C. Volant[4], and J. P. Wieleczko[1]

(INDRA collaboration)

[1]*GANIL, CEA et IN2P3-CNRS, B.P. 55027, F-14076 Caen Cedex 05, France*
[2]*LPC, IN2P3-CNRS, ISMRA et Université, F-14050 Caen Cedex, France*
[3]*Institut de Physique Nucléaire, IN2P3-CNRS, F-91406 Orsay Cedex, France*
[4]*DAPNIA/SPhN, CEA/Saclay, F-91191 Gif-sur-Yvette Cedex, France*
[5]*Dip. di Scienze Fisiche e Sez. INFN, Università di Napoli "Federico II", Napoli, Italy*
[6]*Institut de Physique Nucléaire, IN2P3-CNRS et Université, F-69622 Villeurbanne Cedex, France*
[7]*Conservatoire Nationale des Arts et Métiers, Paris, France*
[8]*Laboratorio Nazionale del Sud, Via S. Sofia 44, I-95123 Catania, Italy*
[9]*Nat. Inst. For Physics and Nuclear Engineering, Bucharest-Magurele, Romania*
[10]*Laboratoire de Physique Nucléaire, Université Laval, Québec, Canada*
[#]Deceased

**Abstract**

The universal theory of order parameter fluctuations ($\Delta$-scaling laws) is applied to a wide range of intermediate energy heavy-ion collision data obtained with INDRA. This systematic study confirms that the observed fragment production is compatible with aggregation scenarios for in- or out-of-equilibrium continuous phase transitions, while not showing any sign of critical behaviour or phase coexistence. We stress the importance of the methodology employed in order to gain further insight into the mechanism(s) responsible.

## 1. Introduction

The search for signatures of the predicted liquid-gas phase transition of nuclear matter in hadron- and nucleus-nucleus collisions has been the focus of experimental and theoretical activity since the early 1980's [1]. Recently, several new analyses of multifragmentation data have given convincing arguments to suggest that the quest may be nearly at an end [2-4], heralding the dawn of a new age of 'quantitative nuclear thermodynamics'.

However it is only with great difficulty and care that one may extract the nuclear matter phase diagram from collisions that produce small, finite, 'open', and strongly fluctuating systems, whose relaxation and decay times are of the order of the duration of the collision. In such a situation the validity of an approach based on standard (i.e. Gibbs') equilibrium statistical physics and

---

thermodynamics is highly questionable. It requires a solid theoretical grounding for the ergodic hypothesis used to extract information on (time-)equilibrated excited nuclear matter from the study of ensembles of non-identically prepared and not-necessarily-equilibrated systems (i.e. the statistical ensembles generated by the analysis' event selection). In the absence of advances in this direction, such approaches can at best provide only an effective description of the physics of nuclear fragmentation.

We will attempt to show here what is the most that can be said about a possible phase transition in intermediate energy heavy-ion collisions in the least hypothesis-dependent way possible. For this we have applied the universal fluctuations theory of Botet and Ploszajczak [5] to a large set of data obtained with the INDRA multidetector. Details of the experimental set-up can be found in [6]. We will first describe the analysis protocol. General features of the evolution of fragment production with beam energy will be shown. The analysis will then be applied to data, including for the first time a systematic study of the scaling dependence on system mass, and we will discuss the interpretation of these results.

## 2. $\Delta$-scaling analysis protocol

Universal scaling laws of fluctuations (the $\Delta$-scaling laws) can be derived for equilibrium and off-equilibrium systems when combined with the finite-size scaling analysis. In any system in which the second-order critical behaviour can be identified, the relation between order parameter, criticality and scaling law of fluctuations has been established and the relation between the scaling function and the critical exponents has been found. Details can be found in [5].

Experimental observables that may be related to a critical order parameter can be identified through their $\Delta$-scaling behaviour. The $\Delta$-scaling is observed when two or more probability distributions $P_N[m]$ of the observable $m$ collapse onto a single scaling curve $\Phi(z_{(\Delta)})$ when plotted in terms of the scaling variables:

$$\langle m \rangle^{\Delta} P_N[m] \equiv \Phi(z_{(\Delta)}) \equiv \Phi\left(\frac{m - m^*}{\langle m \rangle^{\Delta}}\right) \qquad (1)$$

where $m^*$ is the most probable value of $m$ and $1/2 \leq \Delta \leq 1$. The scaling law Eq. (1) with $\Delta = 1/2$ is in some sense trivial and corresponds to systems with short-range correlations, or to observables which are not related to an order parameter. For $m$ to be an order parameter it must exhibit a change of $\Delta$-scaling regime when some suitable control parameter (e.g. available energy, temperature) is varied. The limiting value $\Delta = 1$ corresponds to the maximum possible fluctuation in a finite system.

There are two generic families of fragment production scenarios for which the second-order phase transition has been identified, with two different order parameters. These are the fragment multiplicity (*fragmentation scenarios*, e.g. fragmentation-inactivation-binary model[7]) and the size of the largest fragment (*aggregation scenarios*, e.g. percolation model).

It was shown in [8] for central intermediate energy heavy-ion collisions that the fragment multiplicity is not an order parameter (trivial $\Delta$-scaling behaviour). On the other hand the size of the largest fragment in each event, $Z_{max}$, has fluctuations which exhibit non-trivial scaling. In this contribution we will continue and extend the study of this observable's Δ-scaling properties.

### 3. Experimental data for central Xe+Sn collisions from 25 to 150 AMeV

Collisions of $^{129}$Xe+$^{nat}$Sn nuclei were studied with INDRA at the GANIL and GSI accelerator facilities in order to make a systematic study of the evolution of fragment production using the same experimental apparatus. Experimental details can be found in [9,10]. We used a global variable, the total transverse energy of light charged particles (isotopes of H and He), $E_{t12}$, in order to estimate a geometrical impact parameter according to [11]. We defined cuts for "central collisions" such that the measured cross-section for $E_{t12}$ values higher than the cut-off was either 10% ($b_{red}$<0.3) or 1% ($b_{red}$<0.1) of the total measured cross-section for all events accepted by the data acquisition trigger (at least 4 fired detector modules). The two data samples obtained in this way are qualitatively the same and are compatible with a maximal geometrical overlap between projectile and target. By making such a selection our aim is simply to study those collisions where the interaction between projectile and target is the most violent possible, without making any hypothesis about what the result of such a collision may be (formation of an equilibrated "source" etc.). An additional condition on the total detected charge (>80%) was used for the Δ-scaling analysis in order to ensure that the largest *detected* fragment is very probably the largest *produced* fragment.

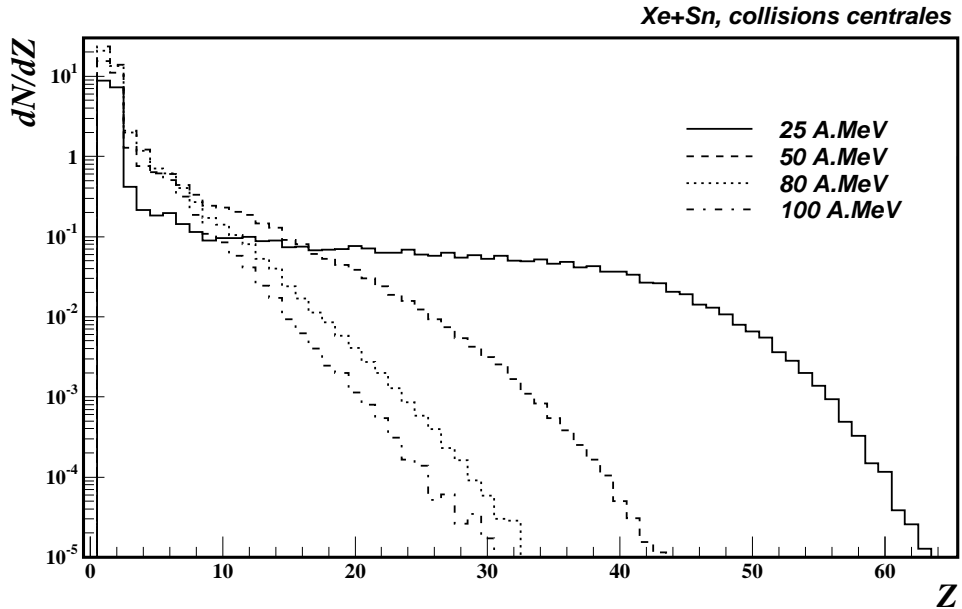

*Figure 1. Central collisions ($b_{red}$<0.3) of Xe+Sn measured with INDRA: evolution of reaction product charge distributions with bombarding energy.*

Figure 1 shows how the number and size (charge) of reaction products depend on bombarding energy. The overall charge distribution of products evolves from a very wide distribution at 25AMeV, for which residues heavier than projectile or target are seen to survive, towards an exponential form indicating complete fragmentation of both projectile and target at 100AMeV. Mean centre of mass kinetic energies of fragments as a function of their charge (figure 2) show, at the lowest incident energy (25AMeV), the survival of slow-moving heavy residues with mean energies smaller than those of the intermediate mass fragments. This suggests considerable stopping and some kind of (incomplete) fusion-like character for central collisions at this energy.

At higher energies (≥50AMeV) one no longer observes any distinguishing behaviour of the heaviest fragments as regards their kinetic energy, and the mean kinetic energy of all fragments increases monotonously with their charge (mass), indicating some degree of "transparency" at these energies, in agreement with AMD simulations [10,12]. With increasing bombarding energy, fragment kinetic energies increase more steeply with Z, while the size of the largest fragment surviving the collision decreases as already pointed out above.

The fragment production in these data does not present any obvious sign of an abrupt phase transition: there is no discontinuity, for example, in the fragment multiplicity as a function of beam energy, and fragment sizes evolve continuously from the "evaporation residue" regime towards complete disassembly. However, some potentially interesting behaviour may appear if we look at higher moments of observables rather than simple mean values (first moments). We studied the normalised (or reduced) fluctuations, $\sigma/<m>$, as a function of incident energy. As regards the multiplicities of LCP or of IMF, no particularly striking behaviour of their fluctuations comes to light. On the other hand the normalised fluctuations of $Z_{max}$ increase rapidly at low energy, reach a sharply defined maximum around 45-50AMeV, and then saturate. As we will see below, this behaviour has a simple interpretation in the framework of the Δ-scaling analysis.

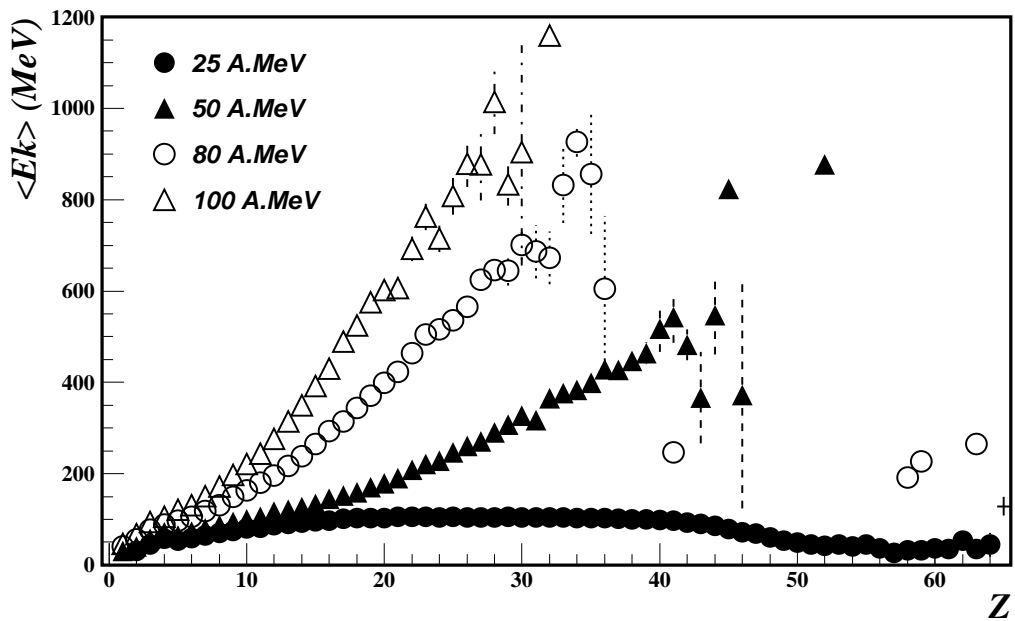

*Figure 2. Mean c.m. kinetic energies of charged reaction products as a function of their charge, Z, for central ($b_{red}$<0.3) Xe+Sn collisions at 25, 50, 80, and 100AMeV.*

# 4. Δ-scaling analysis

## *4.1 Energy dependence for fixed system mass (central collisions)*

Figure 3 shows data for central collisions of Xe+Sn from 25 to 150AMeV, expressed in terms of the $\Delta=1/2$ and $\Delta=1$ scaling laws. The distributions of the charge of the heaviest fragment per event collapse to a single curve with $\Delta=1$ for bombarding energies 45AMeV and above. At lower energies a $\Delta=1/2$ scaling law is obeyed. We have checked that, as in [8] where a different selection of central collisions was used [13], the fragment multiplicity distributions obey the "trivial" $\Delta=1/2$ scaling law whatever the incident energy. On the other hand, the change of scaling regime for the size of the largest fragment is here observed at a higher incident energy than in [8] (45AMeV rather than 32AMeV). This is due to the event selection we used here. The requirement of large LCP transverse kinetic energies reduces the energy available for fragment production and so a higher incident energy is necessary in order to achieve the same degree of fragmentation, compared to the previous selection based on the isotropy of fragment momenta.

On a log-log plot of the second and first cumulant moments of the experimental distributions (figure 4), data which obey the Δ-scaling fall on a straight line of slope Δ. Thus the change of scaling regime observed for the Xe+Sn system can clearly be seen as a change of slope at around 45AMeV incident energy. With increasing energy (going from right to left) one passes from the $\Delta=1/2$ "branch" where the fluctuations of the size of the largest fragment grow with the mean value as $\sigma \sim <Z_{max}>^{1/2}$ to the $\Delta=1$ branch with $\sigma \sim <Z_{max}>$ i.e. $\sigma / <Z_{max}> \sim$ constant. The saturation of the reduced fluctuations of $Z_{max}$ observed above corresponds to the transition to the regime of maximal fluctuations.

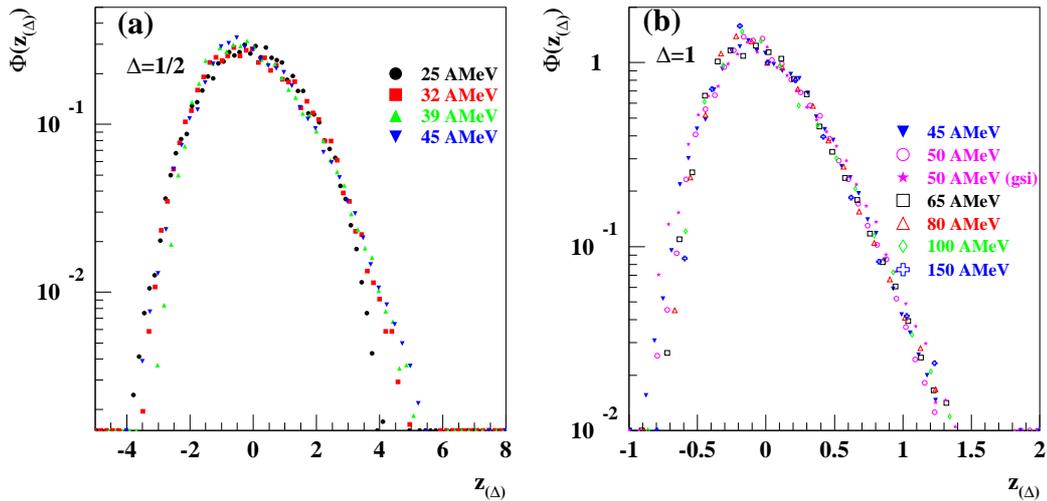

*Figure 3. Probability distributions of the charge of the largest fragment in each event, $Z_{max}$, plotted in the variables of the (a) $\Delta=1/2$ and (b) $\Delta=1$ scaling. Central collisions ($b_{red}<0.1$) of Xe+Sn.*

*4.2 Total mass, energy, and entrance channel dependence for central collisions*

The non-trivial dependence of the Δ-scaling of the largest fragment on incident energy is also observed for a wide range of systems with very different masses. Figure 4 summarises this result but it should be noted that the scaling properties were first determined by examining the scaling functions such as shown in Figure 3. The Xe+Sn (A=248) behaviour is quite closely followed by the lighter (A=116) Ni+Ni system, with two separate branches clearly in evidence. The preliminary analysis for the Au+Au system (A=394), for which only two energy points are shown, seems to confirm the trend of the Xe+Sn data although more points must be included in order to determine whether the lowest energy (40AMeV) really belongs to the Δ=1 scaling branch or not. For the two lightest systems, Ar+KCl (A=73) and Ar+Ni (A=94), all of the available data up to the highest measured beam energy can be described quite well by a Δ=1/2 scaling law although data for higher energies would be necessary in order to check whether or not the last data points show the beginning of a Δ=1 branch (see Figure 4).

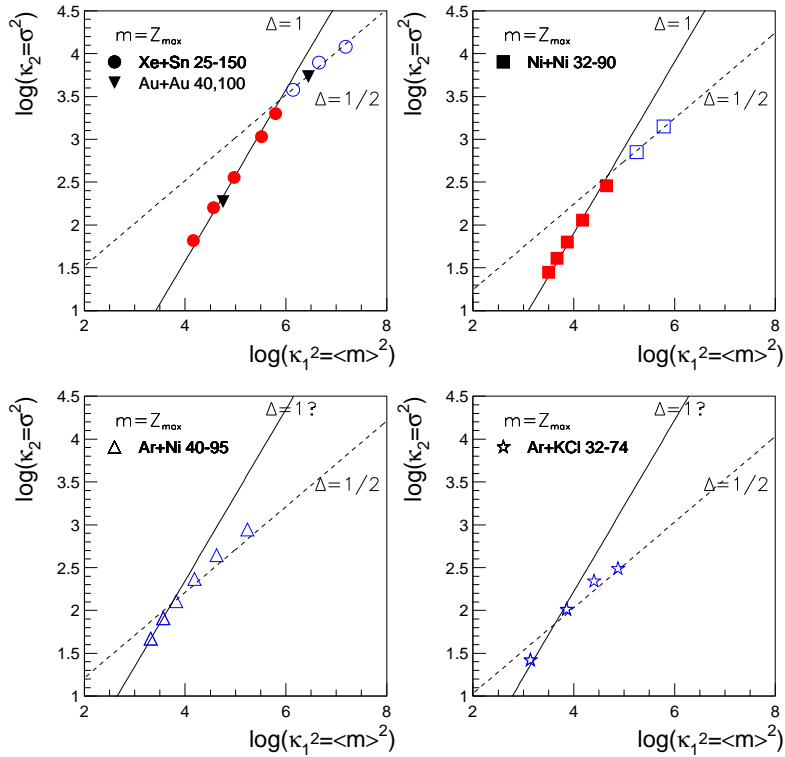

*Figure 4. Compilation of Δ-scaling results concerning the largest fragments in central collisions ($b_{red}$<0.1) of Au+Au, Xe+Sn, Ni+Ni, Ar+Ni and Ar+KCl from 25 to 150AMeV. For each system the lowest bombarding energy (largest mean $Z_{max}$) corresponds to the rightmost point on the plot, the incident energy increases going towards the left. The lines marked Δ=1/2 and Δ=1 are to guide the eye. Open (closed) symbols represent systems obeying a Δ=1/2 (Δ=1) scaling.*

We observe therefore a systematic mass-dependence of the beam (available) energy at which the transition to the maximal fluctuation regime occurs. This transition energy decreases with increasing mass, from ~65AMeV for Ar+KCl (A=73) to ~40AMeV for Au+Au (A=394). There is also possibly some entrance channel dependence, as the asymmetric Ar+Ni system's transition energy does not follow the general trend even when expressed as available (centre of mass) energy.

*4.3 Excitation energy dependence (central collisions)*

In [14] a subset of central collisions of Xe+Sn that is compatible with fragment emission from a single, thermalised source was shown to exhibit a negative branch of its calculated "heat capacity" as a function of the estimated source excitation energy. This is related to a maximum in energy fluctuations at the assumed "freeze-out" stage. The $\Delta$-scaling analysis for exactly the same data sets is shown in Figure 5 (round symbols). The 32AMeV data, sorted into thermal excitation energy bins, shows the same change of scaling regime as was observed in [8] by varying the beam energy. More importantly, the transition takes place at the middle of the excitation energy range, which is near to where the "heat capacity" of [14] exhibits the so-called "second divergence": for the points belonging to the $\Delta=1/2$ branch the "heat capacity" is negative, while for the $\Delta=1$ branch it is positive.

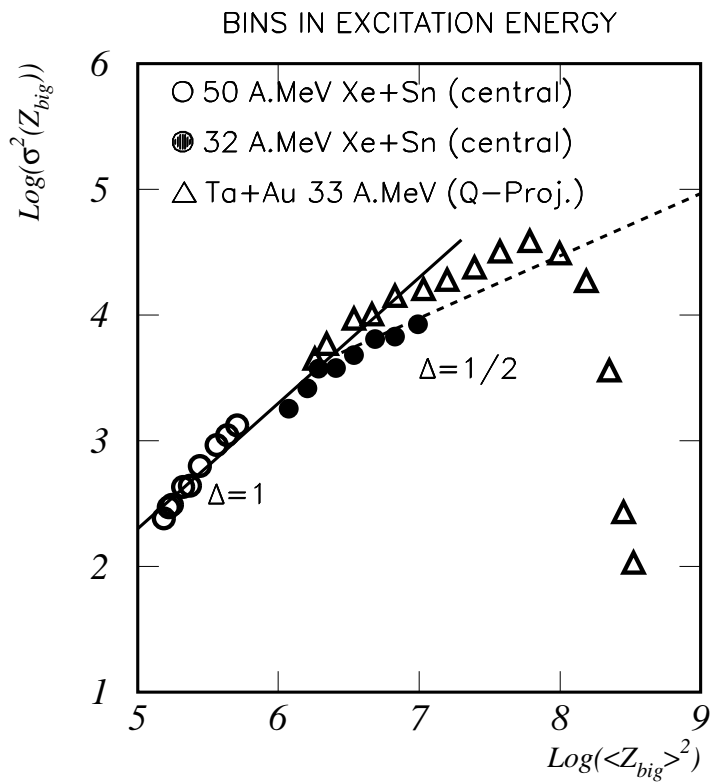

*Figure 5. Log-log plot of the 1st and 2nd cumulant moments of the distribution of the charge of the largest fragment in each event, $Z_{big}$, for events sorted according to estimated source excitation energy (see text). The lines marked $\Delta=1/2$ and $\Delta=1$ are to guide the eye.*

The 50AMeV data are consistent with this: for virtually all excitation energies the "heat capacity" was found to be positive at this beam energy, and the data points show that the fluctuations of the largest fragment obey approximately the $\Delta=1$ scaling law. However it may be remarked that these data actually appear to follow an unphysical $\Delta>1$ scaling never observed for data sorted simply according to beam energy. This is perhaps due to the reconstruction of the thermal excitation energy.

*4.4 Excitation energy dependence (peripheral collisions)*

In [8] a first study of the impact parameter dependence of the $\Delta$-scaling analysis was made by relaxing the conditions used to select central collisions. In Figure 5 tantalum quasi-projectiles from 33AMeV Ta+Au collisions were isolated and sorted according to the estimated excitation energy of the reconstructed sources (see [14] for details of the method). The resulting distribution of excitation energies is far wider than for central collisions (Figure 5, open triangles). Most of the data obey a $\Delta=1/2$ scaling law, with perhaps the beginnings of a transition to the $\Delta=1$ regime for the very highest "excitation energies". It is interesting to observe how fluctuations are rapidly suppressed at the lowest energies (E*< ~3AMeV) where the data for $Z_{max}$ ($Z_{big}$ in Figure 5) no longer obey any meaningful scaling law. This is due to the closing of decay channels that allow the size of the largest fragment to fluctuate (it should be noted that "fission" events have been excluded): below the threshold for charged particle emission the probability distribution of $Z_{max}$ for a given source size is a $\delta$-function.

**5.Interpretation of results**

The universal scaling laws of order parameter fluctuations were derived using only minimal assumptions about the system under study. They were then applied to various model calculations in [15], the results of which provide possible interpretations of the behaviour observed in the data. However one should always remember that the interpretations, unlike the scaling laws, are model-dependent.

The fact that, in all data studied so far, it is the size of the largest fragment and not the fragment multiplicity that exhibits non-trivial scaling behaviour excludes the *fragmentation scenarios* of fragment production and suggests that nuclear multifragmentation most resembles *aggregation scenarios* such as the percolation, Fisher droplet or Smoluchowski gelation models, the latter describing an off-equilibrium process. Indeed it is important to note that as the theory of universal fluctuations is independent of any equilibrium hypothesis the success of its application gives no information whether or not multifragmentation in intermediate energy heavy-ion collisions is an equilibrium (or thermal) process.

The $\Delta=1/2$ and $\Delta=1$ scaling laws are compatible with the *ordered* and *disordered* phases seen, for example, in *sub-* and *supercritical* percolation calculations. These should not be confused with the *liquid* and *gas* phases of a van der Waals' fluid: indeed, a subcritical fluid may be composed of liquid, gas, or a mixture of the two (coexistence), depending on the density and pressure. The scaling behaviour shown in Figure 5 for central Xe+Sn collisions is therefore consistent with the "heat capacity" analysis of [14], if one associates negative capacities with the coexistence region and positive (at energies above the "second divergence") with supercritical regions of the nuclear matter phase diagram. Consistent results are also obtained in a Fisher droplet analysis [16]. On the other hand for coexistence one would expect a bimodal scaling function for $Z_{max}$ [8] which has not been observed here whether events are sorted according to excitation energy ("microcanonical" ensembles) or not.

More generally speaking, our systematic study shows that, for a wide range of colliding systems of different mass, a change in the scaling properties of the fluctuations of the largest fragment occurs which is compatible with the passage from subcritical to supercritical systems in e.g. percolation models. The available (c.m.) energy at which this change of fluctuation regime takes place decreases with system mass, and seems to be between 10 and 16AMeV.

The detailed study of the scaling function, which holds all essential information on the properties of the system, allows us to exclude the possibility of any passage by a critical point of the systems studied in these data. Indeed for critical systems the tail of the scaling function for large positive $z_{(\Delta)}$ (see Eq.1) should fall off faster than a gaussian. In all cases studied here no such behaviour has been found but may yet be revealed by new data taken last year in small steps of bombarding energy.

### 6. Conclusions

Central collisions between nuclei at intermediate energies, and more particularly the associated fragment production, do not exhibit any obvious signs of phenomena related to a phase transition of nuclear matter. Nevertheless, many "signals" predicted to be associated with phase transitions are found in data after the appropriate analysis. The problem is that all of the signals are necessary but not sufficient conditions. Among them, the universal scaling laws proposed by Botet and Ploszajczak are unique by the bareness of their basic suppositions. This makes it far easier to avoid confusion between what has actually been observed (and indeed another advantage is that the analysis only requires experimentally measured quantities) and interpretations based on models and/or additional hypotheses.

The least biased interpretation of the results presented here is that: (a) of the two generic families of fragment production scenarios for which the second-order phase transition has been identified, fragment production in central heavy-ion collisions from 25-150AMeV shows clear affinities with the aggregation scenarios having the size of the largest fragment as order parameter; (b) independent of any hypothesis, the $\Delta$-scaling analysis shows that above a certain energy, and if the system is heavy enough, the size of the heaviest fragment produced in central collisions begins to exhibit fluctuations which are the largest allowed in nature; (c) in this analysis no sign of either critical behaviour or phase coexistence was observed.

Faced with a plethora of necessary but not sufficient signals, we need an appropriate methodology in order to advance our understanding of fragment production in heavy-ion collisions and its relationship with the properties of the nuclear matter phase diagram. It is not enough to show the compatibility of a mechanism with data, all other mechanisms must be shown to be incompatible. This may be impossible to achieve in practice.

Therefore we propose a "minimum-bias" or "bottom-up" protocol inspired by the Δ-scaling analysis. For any proposed signal seen in data, one should look for the "poorest" possible model that reproduces the signal. The hypotheses of this model then constitute the minimum physical ingredients necessary in order to understand the data. This is, in the absence of the necessary theoretical development of a *sufficient* signal, all that we may claim to know with certainty.

*Acknowledgements*

The authors would like to thank the technical staff of GANIL and GSI, as well as the Aladin collaboration, for indispensable assistance in performing the experiments. We also thank M.Ploszajczak for fruitful and stimulating discussions.